# RCRNORM: AN INTEGRATED SYSTEM OF RANDOM-COEFFICIENT HIERARCHICAL REGRESSION MODELS FOR NORMALIZING NANOSTRING NCOUNTER DATA[1]

BY GAOXIANG JIA[*,†], XINLEI WANG[*], QIWEI LI[†], WEI LU[‡],
XIMING TANG[‡], IGNACIO WISTUBA[‡] AND YANG XIE[†]

*Southern Methodist University[*], University of Texas Southwestern
Medical Center[†] and University of Texas[‡]*

Formalin-fixed paraffin-embedded (FFPE) samples have great potential for biomarker discovery, retrospective studies and diagnosis or prognosis of diseases. Their application, however, is hindered by the unsatisfactory performance of traditional gene expression profiling techniques on damaged RNAs. NanoString nCounter platform is well suited for profiling of FFPE samples and measures gene expression with high sensitivity which may greatly facilitate realization of scientific and clinical values of FFPE samples. However, methodological development for normalization, a critical step when analyzing this type of data, is far behind. Existing methods designed for the platform use information from different types of internal controls separately and rely on an overly-simplified assumption that expression of housekeeping genes is constant across samples for global scaling. Thus, these methods are not optimized for the nCounter system, not mentioning that they were not developed for FFPE samples. We construct an integrated system of random-coefficient hierarchical regression models to capture main patterns and characteristics observed from NanoString data of FFPE samples and develop a Bayesian approach to estimate parameters and normalize gene expression across samples. Our method, labeled RCRnorm, incorporates information from all aspects of the experimental design and simultaneously removes biases from various sources. It eliminates the unrealistic assumption on housekeeping genes and offers great interpretability. Furthermore, it is applicable to freshly frozen or like samples that can be generally viewed as a reduced case of FFPE samples. Simulation and applications showed the superior performance of RCRnorm.

**1. Introduction.** Formalin-fixed paraffin-embedded (FFPE) tissue samples are usually collected for diagnostic purposes in clinical routines (Lüder Ripoli et al. (2016)). Unlike freshly frozen (FF) tissue samples that must be frozen instantly after collection and then stored in freezers, FFPE samples can be stored

Received March 2018; revised November 2018.

[1]Supported by National Institutes of Health (R15GM113157 and R15GM131390 to X. W.; P50CA70907 to Y. X. and I.I.W; R01GM115473 to Y. X.) and CPRIT (RP180805 to Y. X.). All correspondence should be addressed to Xinlei Wang. RCRnorm has been implemented by an R package, downloadable from CRAN: https://cran.r-project.org/web/packages/RCRnorm/. Supplementary Data are available at Annals of Applied Statistics Online.

at room temperature and kept for a long time. Due to ease of handling and inexpensive storage (Perlmutter et al. (2004)), numerous FFPE tissue samples have been deposited into tissue banks and pathology laboratories around the world and are readily available (Reis et al. (2011), Lüder Ripoli et al. (2016)). Such samples are often accompanied by well documented patient information, disease status and long-term clinical follow up information. Furthermore, there exist vast archives of specimens from which only FFPE, but no FF, samples can be obtained (e.g., specimens of a deceased patient). Thus, the ubiquity of FFPE samples has made them a highly valuable resource in biomedical studies. In particular FFPE samples have great potential for biomarker discovery which can be critical for disease diagnosis, prognosis and treatment plan selection (Ludwig and Weinstein (2005), Rosenfeld et al. (2008), Xie et al. (2011)).

Despite the advantages of FFPE samples, the formalin fixation process breaks RNA into small pieces with an average size of $\approx$200 nt, and irreversible methylene crosslinks between RNAs and proteins may form that affect enzyme based downstream reactions (Masuda et al. (1999)). The low quality of RNA from FFPE samples hinders reproducibility and sensitivity of assays for quantitatively measuring gene expression levels via microarray experiments or real time polymerase chain reaction (qPCR) which involves enzyme-mediated reverse transcription from mRNA to cDNA (von Ahlfen et al. (2007)). Thus, in order to exploit the vast collection of FFPE samples, robust assays are needed to enable and improve expression profiling in these samples.

In recent years several methods/platforms have been developed for gene expression profiling in FFPE samples either at the genome-wide scale or for a subset of genes. April et al. (2009) developed a whole genome cDNA-mediated annealing, extension, selection and ligation (WG-DASL) assay to perform gene expression profiling in FFPE samples. Iddawela et al. (2016) reported that WG-DASL assays could reliably probe gene expression levels in breast cancer FFPE samples. Abdueva et al. (2010) showed that Affymetrix microarrays could be used to probe gene expression signatures and to perform differential expression analysis with FFPE samples, obtaining results comparable to those from unfixed tissues. Thompson et al. (2014) developed the HTG EdgeSeq chemistry platform that uses RNA extraction-free nuclease protection assay (qNPA), followed by the quantification of RNA molecules by next generation sequencing techniques, such as RNA-seq, to profile microRNA and RNA in FFPE samples.

Unlike whole genome expression profiling discussed above, Paluch et al. (2017) developed targeted RNA-seq that can selectively examine the abundance of immune related genes on archival FFPE samples. Typically, platforms for measuring expression levels of only a subset of genes are called medium-throughput platforms. Compared to the high-throughput (genome-wide) platforms, these platforms often have better technical reproducibility in clinical settings. For medium-throughput platforms, there are usually probes designed for internal controls, such as negative controls, positive controls and housekeeping genes besides the probes for detecting genes of interest. Negative controls target no known sequence and should ideally have zero count; positive controls added to the reaction system have known amounts of RNA targets, and housekeeping genes maintain basic cell functions with expression levels that minimally fluctuate across different individuals compared with other genes (Waggott et al. (2012)). These internal controls can provide information useful in adjusting for unwanted biological and technical effects that can mask the signal of interest.

Among the medium-throughput platforms the highly-multiplexed NanoString nCounter is the most popular (Geiss et al. (2008)); it can effectively detect up to 800 genes in a single tube in one run which bridges the gap between genome-wide expression profiling by microarray or RNA-seq and targeted profiling by qPCR (Kulkarni (2011)). More importantly, the nCounter platform is a Clinical Laboratory Improvement Amendments (CLIA) certifiable assay (Friedman (1997)), meaning it can be translated into clinical settings.

Due to its importance in medium-throughput profiling, several analysis methods, including NanoStringNorm, NAPPA and NanoStringDiff, have been developed for the NanoString nCounter platform to normalize and extract gene expression levels from different samples. These algorithms are mainly focused on removing noise from each of the following three sources with the use of one specific type of internal controls: (1) lane-by-lane noise, resulting from variation in experimental conditions (such as humidity, temperature, etc.) between reaction systems, is estimated and removed by using information from positive controls; (2) background noise, introduced by nonspecific binding of the probes, is estimated and removed using negative controls; and (3) variation in sample loading amounts or difference in RNA degradation levels is evaluated using housekeeping genes (Waggott et al. (2012), Wang et al. (2016)).

Specifically, NanoStringNorm is an R package that implements a normalization protocol recommended by the manufacturer's guideline (Waggott et al. (2012)). First, lane-by-lane variation is removed by scaling the samples with a factor that makes summary statistics of positive control counts (e.g., mean, median or geometric mean) equal across samples. Then, background correction is performed by subtracting the read count with a statistic representing the background noise, like the mean or maximum count of negative controls. Finally, the loading variation is adjusted by a factor calculated from housekeeping genes in the same way as in the first step. It is clear, here, that NanoStringNorm performs normalization in an ad hoc way without any rigorous statistical model involved. To the best of our knowledge, NAPPA is perhaps the algorithm most commonly used by researchers to normalize NanoString data. This algorithm adjusts the background noise with a truncated Poisson distribution and corrects the loading variation by fitting a sigmoidal curve while normalizing the lane-by-lane variation similarly as in NanoStringNorm. NanoStringDiff is originally designed for identifying differentially expressed genes based on the NanoString nCounter platform but can be easily adapted for the purpose of normalization (Wang et al. (2016)). Nano-StringDiff fits a generalized linear model to the data, from which three factors are extracted from positive controls, negative controls and housekeeping genes to adjust for lane-by-lane variation, background noise and variation in the amount of input sample respectively.

Although the three methods are designed or can be used to normalize Nano-String nCounter data, no meticulous research has been conducted to study the characteristics of this type of data obtained from FFPE samples, and no simulation has been carried out to evaluate their performance in normalizing such FFPE data. Furthermore, information provided by different types of internal controls is intermingled. For example, although positive controls are designed to measure noise from varying experimental conditions, read counts from negative controls can also provide useful information about this type of noise. Current normalization methods ignore this fact and cannot make the best use of data. In addition all current algorithms use housekeeping genes by assuming that their expression levels are constant over different samples (i.e., zero variance), which may not be the case, since biologists generally define housekeeping genes as those that do not vary much across different tissues (Eisenberg and Levanon (2013)). Thus, advanced statistical modeling of FFPE data based on an integrated understanding of the nCounter system without restrictive model assumptions is needed to boost its application in clinical and academic research. While our focus is on FFPE samples, the resulting method can be still applied to FF or like samples. This is because FF samples may be viewed as a reduced case of FFPE samples (i.e., zero or very low degradation levels in FF compared to much higher and more diverse degradation levels in FFPE).

We begin the paper by exploring key features of the NanoString nCounter data from FFPE samples. We then construct an integrated system of random-coefficient hierarchical regression models for modeling log-transformed read counts from the different types of probes in the nCounter system. Our computational strategy is based on a Bayesian approach. The proposed method is labeled by RCRnorm, where "RCR" stands for **r**andom-**c**oefficient hierarchical **r**egression and "norm" stands for normalization. We present a formal simulation study conducted to evaluate the performance of RCRnorm in comparison to several existing normalization methods as well as to examine its robustness to deviations from key model assumptions. Real data applications are provided as well to illustrate the proposed Bayesian approach. Finally, the paper is concluded with a brief summary and some in-depth discussions.

## 2. Motivating example.

2.1. *Data description.* The data that motivate our research are from a published study by Xie et al. (2017) which aims to validate a 12-gene signature for predicting adjuvant chemotherapy (ACT) response in lung cancer. A gene signature is a subset of genes, selected from all human genes (more than 20,000), that can be used for diagnosis or prognosis of diseases such as cancer (Ziober et al. (2006), Chen et al. (2007)). Typically, a gene signature is identified via variable/model selection techniques with each gene's expression measurement corresponding to a variable.

The 12-gene signature was developed from FF samples to predict who, among lung cancer patients, would benefit from ACT so that patients that are unlikely to benefit from ACT can avoid adverse effects of unnecessary treatment (Tang et al. (2013)). As mentioned before, FFPE samples are widespread. FF samples, however, are not readily available for clinical applications, due to reasons including: (i) easy contamination by pathogenic germs, (ii) rapid deterioration in room temperature, and (iii) much higher storage cost for frozen specimens than room temperature specimens (Stefan, Michael and Werner (2010)). Thus, it is important to validate the performance of the signature on FFPE samples so that a clinical applicable assay can be developed based on the nCounter platform.

The dataset contains gene expression levels measured by the nCounter platform on paired FF and FFPE samples from 30 patients. The goal of the study is to verify that each gene's expression levels of the 30 patients from FFPE samples are well correlated with those from paired FF samples so that the statistical model based on the 12-gene signature derived from FF samples can be applied to FFPE samples as well. Although this signature only contains 12 genes, 87 total genes were measured in the dataset.

Table S1 in Supplementary Material (Jia et al. (2019)) shows the data structure derived by combining raw data files for different patient samples, where each row represents a probe, and each column, except for the first two, represents a sample. The first column labeled "CodeClass" indicates the probe type—negative controls, positive controls, housekeeping or regular genes. The second column contains unique probe names. Generally, there are six positive controls (i.e., $P = 6$) in the code set, but the number of negative controls $N$ and the number of housekeeping genes $H$ can vary. The name of each negative or positive control contains a pair of parentheses within which there is a number indicating the concentration amount of RNA added to the system that is targeted by that control. For the six positive controls the RNA amount is 0.125, 0.5, 2, 8, 32 and 128 fM respectively, while, for all negative controls, it is zero since there is no known RNA transcript that can be targeted by the probes. All the other columns in Table S1 contain (transformed) read counts from individual samples. As will be detailed in Section 3, each (transformed) count is denoted by $Y$ with a superscript representing the code-class affiliation, the first subscript denoting the patient ID and the second subscript denoting the probe ID in that code class.

In the study, the (paired) data involve two tables in the form of Table S1, one for FF samples and the other for FFPE samples from the same set of patients. There are eight negative controls, seven housekeeping genes and 87 regular genes besides six positive controls in the data.

Data generated by the nCounter system have to be normalized to account for sample preparation variation, sample content variation and background noise, etc., before they can be used to quantify gene expression and conduct any downstream statistical analysis. Here, the availability of data from FFPE samples would allow us to explore major characteristics of such data and examine key assumptions/hypotheses about the mean structure of the data when developing a new normalization method that aims to improve existing ones. Meanwhile, the availability of data from paired FF samples would enable us to quantitatively assess and compare the performance of any normalization methods developed for the nCounter system. Due to the lack of ground truth, it is generally difficult to compare the performance of different normalization methods on real data. Nevertheless, the data from paired FF samples, after normalization to remove technical effects, can be used to provide a surrogate of the truth. This is because FF tissues are known to maintain RNA very well (much lower degradation of RNA and no methylene crosslink between RNA and proteins) and thus are considered as a gold standard for most molecular assays (Solassol et al. (2011)).

2.2. *Exploratory analysis.* To ensure that data resulting from an nCounter gene expression experiment is of adequate quality to be used in subsequent analysis, we apply quality control procedures according to the NanoString guidelines (2017). Among the 30 patients' FFPE samples with 87 regular genes, two patients and four genes were removed for their compromised data quality because they have mean read counts lower than the maximum count of negative controls. An interesting fact is that the two samples discarded are the oldest among the 30 FFPE samples and were collected before the year of 2000, supporting the notion that storage time is a key factor that influences RNA quality of FFPE samples (von Ahlfen et al. (2007)).

Figure S1 in Supplementary Material (Jia et al. (2019)) explores the mean-variance relationship for the NanoString data of FFPE samples using positive control probes. Among commonly used strategies (e.g., Poisson and negative binomial distributions, certain transformations), there seems to be no perfect solution to model or remove such a relationship, as detailed in Section S1.1 of Supplementary Material. In this paper we apply the log transformation to the raw counts, based on the following considerations. First of all, the log transformation is suggested by the NanoString manufacturer which designs positive control probes in a way that their concentrations range linearly in the log scale from 128 fM to 0.125 fM. The manufacturer further requires that $R^2$, calculated from the regression between the known log concentrations and observed log counts of these positive controls, be higher than 0.95. Second, the coefficients of variation calculated from positive controls range from 17.1% to 26.1% for raw counts, but they are greatly reduced to the range 2.3%–6.6%. Although no transformation can completely remove the mean-variance relationship, the log transformation does help stabilize the variance (relative to the mean level). Last, but not least, the log transformation was applied because the mathematical theory of count distributions is less tractable than that of the normal distribution, as mentioned in Law et al. (2014). In the context of a complex system of (hierarchical) models, as will be detailed in Section 3, we prefer modeling transformed data under a linear regression framework over raw counts under a generalized linear model (GLM) framework in order to achieve analytical tractability of conditional posterior distributions that greatly facilitates Bayesian computation. To avoid $-\infty$ arising from zero counts, we add one to the observed counts before applying the logarithm.

The empirical distributions of the log10 transformed gene read counts of FFPE vs FF samples are showed in Figures 1(a) and (b) respectively, in which each (density) curve, plotted using log read counts of housekeeping and regular genes, corresponds to a patient sample. It is obvious that the locations of the distributions of FFPE samples vary more dramatically than those of FF samples. This indicates the existence of heterogeneity in RNA degradation and fragmentation levels among the 28 FFPE tissue samples, contributing to individual sample effects in transcript abundance. This should be modeled, whenever possible, to enable comparison of gene expression levels between patients after removal of such technical artifacts. Figure 1(c) plots log read counts of six positive control probes vs. patient index for FFPE samples. We find that the zig-zag patterns for the six probes are extremely similar, strongly indicating the existence of the lane effects.

Given a sample $i$, one would expect that the log read count (say $Y_{ij}$) of any probe $j$ has a monotonically increasing relationship with the corresponding RNA amount (say $R_{ij}$). Using positive controls whose RNA amounts are known and fixed for all $i$ (i.e., $R_{ij} \equiv R_j$ and $R_j$'s are known), we show two violin plots based on FFPE samples in Figure 1(d), one for the 28 patient-wise Pearson correlation coefficients between $X_{ij} \equiv \log R_{ij}$ and $Y_{ij}$ and the other for those between $R_{ij}$ and $Y_{ij}$. As we anticipate according to the manufacturer's guidelines, the coefficients using log RNA amounts are very close to 1, while those using original RNA amounts are much lower. Thus, a linear relationship between $X_{ij}$ and $Y_{ij}$ may capture the underlying pattern well, that is, for each sample $i$,

(1) $$E(Y_{ij}|X_{ij} = x, a_i, b_i) = a_i + b_i x,$$

where $a_i$ and $b_i$ are sample-specific regression coefficients.

Figures 1(e) and (f) show the empirical densities of $a_i$'s and $b_i$'s, all estimated using FFPE data from positive controls. Apparently, the simplifying assumption $a_i \equiv a$ or $b_i \equiv b$ is not appropriate here. Comparing (e) with (f), we can see that the two distributions are different in both location and spread. The Shapiro–Wilk test (Shapiro and Wilk (1965)) suggests no gross departure from normality at the significance level 0.05 for either distribution. Thus, it is plausible to assume that $a_i$'s and $b_i$'s are random and follow two separate normal distributions.

For every housekeeping or regular gene, the RNA amount $R_{ij}$ reflects gene $j$'s expression abundance in sample $i$, whose value is unknown. But for negative controls, $R_{ij} \equiv 0$ so that $X_{ij} = -\infty$ which is ill defined. To solve this issue, we add a small positive number $\delta$ so that $X_{ij} = \log \delta$ instead, and (1) holds for negative controls as well. Note that $\delta$ can be interpreted as the mean nonspecific binding level due to background noise. Further, for the housekeeping and regular genes, the additive effect of $\delta$ on the RNA amount $R_{ij}$ can be ignored for mathematical convenience since $\delta$ is very small. As will be shown in Section 6, the estimate $\hat{\delta}$ is 0.013 and 0.018 in our two real data sets respectively. Both $\delta$ and $R_{ij}$'s are estimable. The intuition is that with the information from positive controls, we can pin down $(a_i, b_i)$ for each sample so that with observed counts from negative controls, we can estimate $\delta$ and, with observed counts from housekeeping or regular genes, we can estimate $R_{ij}$'s.

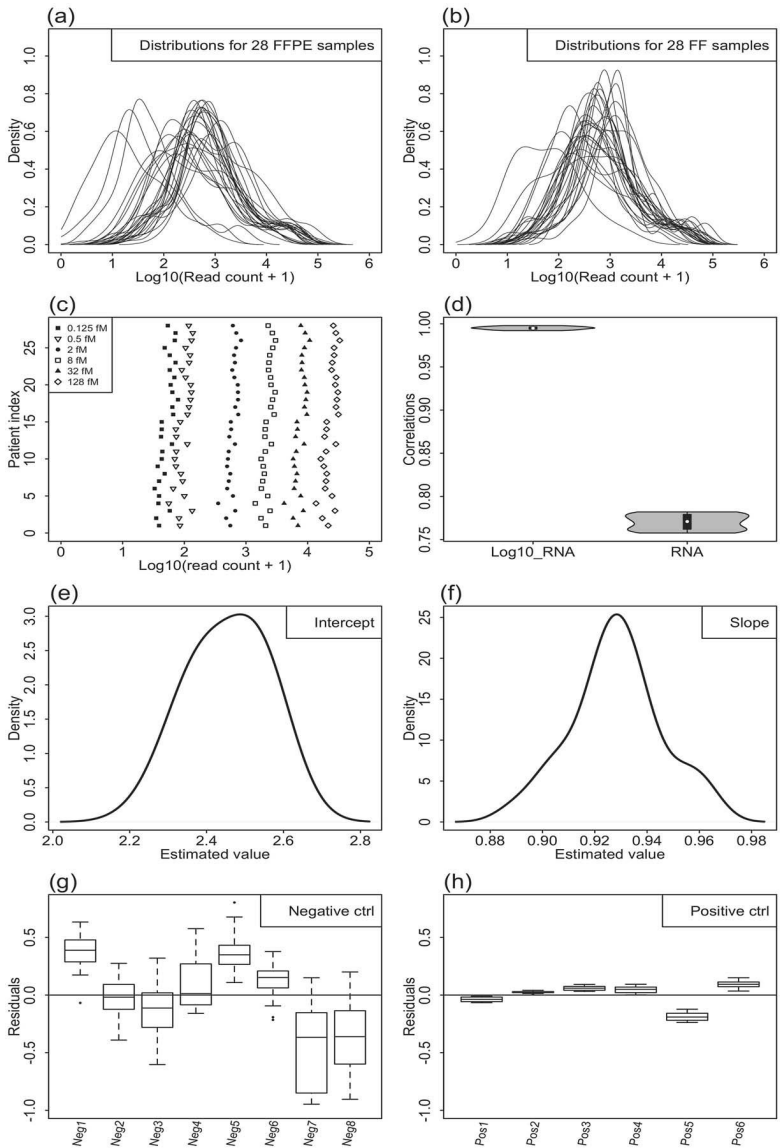

FIG. 1. *Exploratory analysis of lung cancer data from Xie et al. (2017). Panels* (a) *and* (b) *show empirical densities of log read counts based on housekeeping and regular genes for the* 28 *FFPE and FF samples respectively. For FFPE samples panel* (c) *plots log read counts of six positive controls (with different known RNA concentration amounts) vs. patient index;* (d) *shows two violin plots using data from positive controls, one for patient-wise Pearson correlation coefficients between log RNA amount and log read count, and the other for those between RNA amount and log read count;* (e) *and* (f) *show empirical densities of patient-wise intercepts and slopes respectively, estimated using data from positive controls and* (g) *and* (h) *show boxplots of residuals for the eight negative controls and six positive controls respectively, from fitting the linear trend* (1) *per patient, where each boxplot contains residuals from* 28 *patients for a control probe.*

We use $\frac{1}{I \times N} \sum_{i=1} \sum_{j \in \mathcal{J}^-} (Y_{ij} - \hat{a}_i) / \hat{b}_i$ to obtain a rough estimate of $\log \delta$ for FFPE samples, where $\mathcal{J}^-$ denotes the index set of negative controls, $\hat{a}_i$ and $\hat{b}_i$ are estimated using data from positive controls as before and $N$ is the total number of negative controls. We then compute the residuals, that is, deviations from the linear pattern (1), for each positive and negative control, and their boxplots are shown in Figure 1(g) and (h). Two interesting observations can be made here which will be useful for the model construction in Section 3. First, negative controls tend to have much larger deviations than positive controls, and their distributions tend to have much larger variability (hence wider spreads). Second, for each individual probe, the residuals are not randomly distributed around zero. All the boxplots for positive controls are entirely above or below zero, and most boxplots for negative controls have 75% residuals or more above or below zero, indicating residuals are clustered by probes. This can be referred to as probe effects that have been well documented in microarray literature (Li and Wong (2001), Irizarry et al. (2003)).

## 3. Proposed data model based on RCR.
Let $i$ index (FFPE) patient samples, $p$ index positive controls, $n$ index negative controls, $h$ index housekeeping genes and $r$ index regular genes, for $i = 1, \ldots, I$, $p = 1, \ldots, P$, $n = 1, \ldots, N$, $h = 1, \ldots, H$ and $r = 1, \ldots, R$, where $I$ is the number of patients, $P$, $N$, $H$ and $R$ are the (prespecified) number of positive controls, negative controls, housekeeping genes and regular genes respectively, in the NanoString nCounter platform.

Motivated by the analysis in Section 2, we set up a system of (hierarchical) linear regression models with random coefficients for the four different types of probes in which the general linear relationship between the observed log read count and log RNA amount (either known or unknown) is assumed regardless of the probe type and except for the observed log read counts; all the random components of the system are assumed to be independent. We begin with the model for the positive control class, given below:

$$(2) \qquad Y_{ip}^+ = a_i + b_i X_p^+ + d_p^+ + e_{ip}^+,$$

where $Y_{ip}^+$ is the logarithm of read count plus 1 of the $p$th positive control from the $i$th sample, $X_p^+$ represents the logarithm of the known RNA input amount (unit:

$fM$) in the reaction system and the superscript "+" indicates the membership of the positive control class. The $a_i$ and $b_i$ are the sample-specific random intercept and slope which may reflect the lane-by-lane variation. According to Figure 1(e)–(f), we may assume that $a_i$'s and $b_i$'s be independent and identically distributed normal variables respectively, $a_i \overset{\text{i.i.d.}}{\sim} N(\mu_a, \sigma_a^2)$ and $b_i \overset{\text{i.i.d.}}{\sim} N(\mu_b, \sigma_b^2)$. Further, $d_p^+$ represents the probe-specific systematic deviation from the linear pattern (1) (see Figure 1(g)), and we assume $d_p^+ \sim N(0, \sigma_d^2)$. Finally, $e_{ip}^+ \sim N(0, \sigma_e^2)$ is the random error term which reflects the remaining variability of the log-observed count after taking into account the linear trend and the probe-specific deviation.

For the negative control class the model is given by

$$(3) \qquad Y_{in}^- = a_i + b_i c + d_n^- + e_{in}^-,$$

where $Y_{in}^-$ is the logarithm of read count plus 1 of the $n$th negative control from the $i$th sample, $c \equiv \log \delta$ is an unknown constant, the superscript "−" indicates the membership of the negative control class and the other terms are defined similarly, as in (2). As shown in Figure 1(g), the distributions of deviations (from the main linear pattern) for positive controls are very different from those for negative controls. From the centers (i.e., middle horizontal bars) of the boxplots, we can see $d_n^-$'s vary more than $d_p^+$'s and, from the widths of the boxplots, we can see $e_{in}^-$'s vary much more than $e_{ip}^+$'s. Thus, we have to assume $d_n^- \sim N(0, \sigma_{d-}^2)$ and $e_{in}^- \sim N(0, \sigma_{e-}^2)$, where the data suggest that $\sigma_{d-}^2 > \sigma_d^2$ and $\sigma_{e-}^2 > \sigma_e^2$.

For the housekeeping gene class the model is given by

$$(4) \qquad Y_{ih}^* = a_i + b_i X_{ih}^* + d_h^* + e_{ih}^*,$$

where $X_{ih}^*$ is the unknown log RNA amount of the $h$th housekeeping gene from sample $i$, the superscript "∗" indicates the membership of the housekeeping gene class and the other terms are defined similarly, as before. Unlike positive or negative controls, $X_{ih}^*$ in (4) is random by nature rather than being constant, which can be decomposed into a random term $\kappa_{ih}^*$ and a fixed term $\phi_i$, that is, $X_{ih}^* = \phi_i + \kappa_{ih}^*$. Here, $\phi_i$ is a constant that reflects the individual effect of sample $i$ in transcript abundance (e.g., patient-to-patient variation, variation in RNA degradation and fragmentation levels of FFPE tissues, variation in the amount of input sample material, etc.), satisfying $\sum_{i=1}^I \phi_i = 0$, and $\kappa_{ih}^* \sim N(\lambda_h^*, \sigma_{\kappa*}^2)$ reflects the remaining expression abundance after adjusting for the sample effect. Note that $E(\bar{X}_{\cdot h}^*) = \lambda_h^*$, where $\bar{X}_{\cdot h}^* = \sum_{i=1}^I X_{ih}^*/I$ and $\lambda_h^*$ is the gene-specific mean of the log RNA amount. Besides, individual sample effects, $\phi_i$'s, and gene effects, $\lambda_h^*$'s, are both modeled as fixed effects instead of random effects. This is because, for a specific sample, we are interested in recovering $\kappa_{ih}^*$ from $X_{ih}^*$, rather than inferring the marginal distributions of $\phi_i$'s and $\lambda_h^*$'s.

For the regular gene class the RNA amounts in different samples are unknown, too. So, the model is set to be the same as that for the housekeeping gene class but with a difference probe index $r$ and no superscript (for notational brevity),

$$(5) \qquad Y_{ir} = a_i + b_i X_{ir} + d_r + e_{ir},$$

where $X_{ir} = \phi_i + \kappa_{ir}$ is the unknown log RNA amount of the $r$th regular gene from sample $i$, and the definitions of $\kappa_{ir}$, $d_r$ and $e_{ir}$ are self-evident. Correspondingly, we assume $\kappa_{ir} \sim N(\lambda_r, \sigma_\kappa^2)$. Note that two separate variances, $\sigma_{\kappa*}^2$ and $\sigma_\kappa^2$, are needed for the housekeeping and regular genes respectively. This is because expression levels of housekeeping genes are believed to be more stable across samples, so one would expect $\sigma_{\kappa*}^2 < \sigma_\kappa^2$.

In the reaction system of the nCounter platform, negative controls have no known target, and all detected binding signals should be from nonspecific binding while positive controls, housekeeping and regular genes all have known targets, and so their working mechanisms may be similar. Thus, we assume $d_p^+, d_h^*, d_r \sim N(0, \sigma_d^2)$ and $e_{ip}^+, e_{ih}^*, e_{ir} \sim N(0, \sigma_e^2)$. We comment that for the housekeeping and regular genes, (4) and (5) are both hierarchical. The bottom layer involves a linear regression model with random coefficients, and the second layer (for the unknown log RNA amount) involves a two-way ANOVA model, where one factor represents the sample-specific effect $\phi_i$ and the other factor represents the gene-specific effect that is related to $\lambda_h^*$ or $\lambda_r$. In addition for all the four classes, since gene-specific deviations from the main linear trend (1) are allowed through $d_p^+$, $d_n^-$, $d_h^*$ and $d_r$, the log read counts of the same probe from different samples (e.g., $Y_{ir}$ and $Y_{i'r}$) are correlated; meanwhile, the log read counts of the different probes from the same sample (e.g., $Y_{ir}$ and $Y_{ir'}$) are correlated too, as they share the same random intercept $a_i$ and slope $b_i$.

## 4. Bayesian approach.

### 4.1. *Full probability model.*
Based on the system of equations (2)–(5), the parameters of our data model include $\mu_a, \sigma_a^2, \mu_b, \sigma_b^2, \sigma_d^2, \sigma_e^2, \sigma_{d-}^2, \sigma_{e-}^2, \{\phi_i\}_{i=1}^{I-1}$, $\{\lambda_h^*\}_{h=1}^H, \sigma_{\kappa*}^2, \{\lambda_r\}_{r=1}^R, \sigma_\kappa^2$ and $c$, among which $\mu_a, \mu_b, c, \phi_i$'s, $\lambda_r$'s and $\lambda_h^*$'s are location parameters, and all others are variance parameters.

We assume all these parameters are a priori independent. Let $\boldsymbol{Y}$ denote all the observed log read count data; let $\boldsymbol{\Theta}$ denote the collection of all latent random variables involved, including $\{a_i\}_{i=1}^I$, $\{b_i\}_{i=1}^I$, $\{d_p^+\}_{p=1}^P$, $\{d_n^-\}_{n=1}^N$, $\{d_h^*\}_{h=1}^H$, $\{d_r\}_{r=1}^R$, $\{\{\kappa_{ih}^*\}_{h=1}^H\}_{i=1}^I$ and $\{\{\kappa_{ir}\}_{r=1}^R\}_{i=1}^I$, and all model parameters. We also use $N(x|\mu, \sigma^2)$ to denote a normal distribution with mean $\mu$ and variance $\sigma^2$ and $\pi(\cdot)$ to denote a general prior distribution. Then, the full probability model is given by

$$p(\boldsymbol{Y}, \boldsymbol{\Theta}) \propto \prod_{i=1}^I \left\{ \prod_{p=1}^P N\big(Y_{ip}^+ | a_i + b_i X_p^+ + d_p^+, \sigma_e^2\big) \right.$$

$$\cdot \prod_{n=1}^N N\big(Y_{in}^- | a_i + b_i c + d_n^-, \sigma_{e-}^2\big)$$

$$\cdot \prod_{h=1}^{H} N\big(Y_{ih}^* \,|\, a_i + b_i(\phi_i + \kappa_{ih}^*) + d_h^*, \sigma_e^2\big)$$

$$\cdot \prod_{r=1}^{R} N\big(Y_{ir} \,|\, a_i + b_i(\phi_i + \kappa_{ir}) + d_r, \sigma_e^2\big)$$

$$\cdot \prod_{h=1}^{H} N\big(\kappa_{ih}^* \,|\, \lambda_h^*, \sigma_{\kappa*}^2\big) \cdot \prod_{r=1}^{R} N\big(\kappa_{ir} \,|\, \lambda_r, \sigma_\kappa^2\big)$$

$$\cdot N\big(a_i \,|\, \mu_a, \sigma_a^2\big) \cdot N\big(b_i \,|\, \mu_b, \sigma_b^2\big)\bigg\}$$

$$\cdot \prod_{p=1}^{P} N\big(d_p^+ \,|\, 0, \sigma_d^2\big) \cdot \prod_{n=1}^{N} N\big(d_n^- \,|\, 0, \sigma_{d-}^2\big)$$

$$\cdot \prod_{h=1}^{H} N\big(d_h^* \,|\, 0, \sigma_d^2\big) \cdot \prod_{r=1}^{R} N\big(d_r \,|\, 0, \sigma_d^2\big)$$

$$\cdot \pi(\mu_a) \cdot \pi(\mu_b) \cdot \pi(c) \cdot \prod_{i=1}^{I} \pi(\phi_i) \cdot \prod_{h=1}^{H} \pi(\lambda_h^*) \cdot \prod_{r=1}^{R} \pi(\lambda_r)$$

$$\cdot \pi(\sigma_a^2) \cdot \pi(\sigma_b^2) \cdot \pi(\sigma_d^2) \cdot \pi(\sigma_e^2) \cdot \pi(\sigma_{d-}^2) \cdot \pi(\sigma_{e-}^2) \cdot \pi(\sigma_{\kappa*}^2) \cdot \pi(\sigma_\kappa^2),$$

where the first eight lines represent the joint likelihood of the observed data and latent random variables, the last two lines represent prior distributions of the location parameters and the last line represents prior distributions of the variance parameters.

4.2. *Prior specification.* For each variance parameter involved, we specify an inverse gamma prior distribution $IG(u, v)$, where $u$ and $v$ are small positive numbers to make the prior very vague and diffuse (e.g., $u = v = 0.01$). The purpose of doing so is to let the data speak for itself when sampling the variances from the joint posterior distribution.

For $\mu_a$ and $\mu_b$ (mean of random intercepts $a_i$'s and mean of random slopes $b_i$'s), we consider normal priors, $\mu_a \sim N(\hat{\mu}_a, m \times \mathrm{se}(\hat{\mu}_a))$ and $\mu_b \sim N(\hat{\mu}_b, m \times \mathrm{se}(\hat{\mu}_b))$. Here, $m$ is a prespecified constant (e.g., 3, 5) to make the prior much more diffuse than what data suggest; $\hat{\mu}_a$ and $\hat{\mu}_b$ are crude estimates of $\mu_a$ and $\mu_b$, and $\mathrm{se}(\hat{\mu}_a)$ and $\mathrm{se}(\hat{\mu}_b)$ are their standard errors. We simply set $\hat{\mu}_a = \sum_{i=1}^{I} \hat{a}_i / I$ and $\hat{\mu}_b = \sum_{i=1}^{I} \hat{b}_i / I$, where $\hat{a}_i$ and $\hat{b}_i$ are the (least square) estimated intercept and slope from fitting $Y_{ip}^+$ vs. $X_p^+$ for each patient $i$. The standard errors can be estimated using jackknife resampling that removes two patient samples at a time (Efron and Stein (1981)).

For any other location parameter (say $\theta$), we use a noninformative uniform distribution, $\theta \sim \text{Uniform}(L_\theta, U_\theta)$, which should provide a sufficiently wide coverage for all plausible values of $\theta$ suggested by data. For the added small value $\delta$ associated with negative controls, we consider the range $(10^{-6}, 10^{-1})$ so that $c \sim \text{Uniform}(-6, -1)$ is specified a priori. For the gene-specific mean of the log RNA amount $\lambda_r$, the lower and upper bounds can be specified by $\text{mean}(\hat{X}_r) \pm m \times \text{sd}(\hat{X}_r)$, where $\hat{X}_r = (\hat{X}_{1r}, \ldots, \hat{X}_{Ir})$, and $\hat{X}_{ir}$ is a crude estimate of the log RNA amount of regular gene $r$ from sample $i$, for example, $\hat{X}_{ir} = (Y_{ir} - \hat{a}_i)/\hat{b}_i$. The bounds for $\lambda_h^*$ can be specified similarly using $\hat{X}_{ih}^* = (Y_{ih}^* - \hat{a}_i)/\hat{b}_i$ for each housekeeping gene. Note that an alternative method to specify a conservative prior range for any of $\lambda_r$'s and $\lambda_h^*$'s is to use the maximum and minimum statistics, especially when we anticipate that the posterior distribution can be skewed. For example, the lower and upper bounds of $\lambda_r$ can be specified by $\min\{\hat{X}_r\} - \Delta_r$ and by $\max\{\hat{X}_r\} + \Delta_r$ where $\Delta_r$ is a prespecified constant that leaves some extra safe room for either bound (e.g., setting $\Delta_r = \text{sd}(\hat{X}_r)$).

Finally, for the sample effect $\phi_i$, the lower and upper bounds can be specified by $\hat{\phi}_i \pm m \times \text{se}(\hat{\phi}_i)$. Here, the rough estimate $\hat{\phi}_i$ and its standard error can be easily obtained using regular genes by running a standard two-way ANOVA model on $\hat{X} \equiv (\hat{X}_r)_{r=1}^R$ with the constraint that the sum of sample-specific effects and the sum of gene-specific effects are zero. Alternatively, they can be estimated nonparametrically: $\hat{\phi}_i = \bar{\hat{X}}_{i\cdot} - \bar{\hat{X}}_{\cdot\cdot}$, where $\bar{\hat{X}}_{i\cdot} = \frac{1}{R}\sum_{r=1}^R \hat{X}_{ir}$ and $\bar{\hat{X}}_{\cdot\cdot} = \frac{1}{IR}\sum_{i=1}^I \sum_{r=1}^R \hat{X}_{ir}$; and $\text{se}(\hat{\phi}_i)$ can be roughly estimated using jackknife resampling that removes two housekeeping genes at a time.

4.3. *Posterior computation and Bayesian inference.* We use Markov Chain Monte Carlo (MCMC) to draw random samples from the joint posterior distribution $p(\boldsymbol{\Theta}|\boldsymbol{Y})$ which is proportional to $p(\boldsymbol{Y}, \boldsymbol{\Theta})$. Standard diagnostic techniques (Gelman et al. (2014)) are used to detect the convergence. One advantage of the proposed method is that the posterior conditionals, as detailed in Section S2 of Supplemental Material (Jia et al. (2019)), are all known distributions for each of which direct sampling can be done. This property allows us to design an efficient Gibbs sampler in which all the involved quantities are drawn sequentially and generated readily without using any built-in sampling algorithm (such as Metropolis–Hastings and Acceptance/Rejection algorithms) that can greatly slow down the computation.

For the purpose of gene expression normalization, we are mainly interested in estimating $\kappa_{ir}$'s for regular genes. For $i = 1, \ldots, I$ and $r = 1, \ldots, R$, let $\kappa_{ir}^{(t)}$ be the posterior draw of $\kappa_{ir}$ in the $t$th iteration of MCMC after the burn-in period, where $t = 1, \ldots, T$, and $T$ is the total number of iterations. Then, we can estimate $\kappa_{ir}$ by $\tilde{\kappa}_{ir} = \sum_{t=1}^T \kappa_{ir}^{(t)}/T$. Similarly, we can obtain a Bayesian estimate of $\kappa_{ih}^*$ for each housekeeping gene.

## 5. Simulation.

5.1. *Settings.* We conducted five simulation studies, labeled I–V, to examine the performance of the proposed method, called RCRnorm, and to compare it with the three existing methods that have been proposed to normalize NanoString nCounter data, NAPPA, NanoStringDiff, and NanoStringNorm, as mentioned in the Introduction. In addition we included RUV as another competitor, although it is not designed for analyzing NanoString nCounter data. The method, proposed by Risso et al. (2014), normalizes high throughput sequencing data by controlling the factors of unwanted variation (e.g., library preparation) estimated from a set of control genes/samples. RUVseq is its related R package, and is publicly available.

In the first study five settings, labeled I-1 to I-5, were simulated based on the data model proposed in Section 3. In I-1 data were generated using parameter values estimated from the FFPE samples in the lung cancer study described in Section 2. Settings I-2–I-5 were modified from I-1 to mimic different real world scenarios:

I-2: The probed genes have larger variability in their expression levels. To simulate this situation, $\sigma_\kappa$ and $\sigma_{\kappa*}$ were increased to three times that of the basic setting in I-1. These two parameters control the signal strength, where a larger value indicates more genes with strong signals, so it is easier to recover underlying expression levels.

I-3: The samples have larger lane-by-lane variation. This scenario mimics a poor control of experimental conditions across different samples or lanes. To do so, we increased $\sigma_a$ and $\sigma_b$ to three times that of the basic setting in I-1.

I-4: The probe library is poorly designed so that probes have larger variability in their affinity to different gene targets. In this scenario we increased $\sigma_d$ and $\sigma_{d-}$ to three times that of the basic setting in I-1.

I-5: Effects of random errors (unexplained variability) were examined in this scenario by increasing $\sigma_e$ and $\sigma_{e-}$ to three times that of the basic setting in I-1.

RCRnorm, as well as the other methods in literature, assumes a common sample effect $\phi_i$ for all genes in a given sample $i$ to account for between-sample variations resulted from loading or RNA degradation of different samples. However, RNA degradation rates are different among genes as they are determined by a myriad of factors. The gene-wise RNA degradation from either internal pathways or environmental conditions is technically difficult to measure and cannot be separated from true gene expression levels with current data using any of the methods. To reflect this uncertainty, we added white noise $\omega_{ir}$ and $\omega_{ih}^*$, generated from $N(0, 0.4^2)$, to $\phi_i$ for regular or housekeeping genes in the basic setting of our second simulation study, namely II-1. Furthermore, considering that the normality assumption for probe-specific effects and random errors may not always hold, we simulated another five settings by modifying setting II-1, labeled II-2 to II-6, using a standard Student's t distribution with three degrees of freedom ($t_3$), which represents a thick-tailed distribution, and a Gamma distribution with shape 2 and rate 1 which represents a right-skewed distribution. In setting II-2 $t_3$ was used to generate the probe effects $\{d_p^+\}_{p=1}^P$, $\{d_n^-\}_{n=1}^N$, $\{d_h^*\}_{h=1}^H$ and $\{d_r\}_{r=1}^R$, and, in II-3, $t_3$ was used to generate the random errors $\{e_{ip}^+\}_{p=1}^P$, $\{e_{in}^-\}_{n=1}^N$, $\{e_{ih}^*\}_{h=1}^H$ and $\{e_{ir}\}_{r=1}^R$. In II-4 and II-5, Ga(2, 1) was used to generate probe effects and random errors respectively. In II-6 $t_3$ was used to generate the probe effects and Ga(2, 1) to generate random errors. Data generated from $t_3$ were then rescaled to have the same variance as in setting II-1, and data generated from Ga(2, 1) were shifted and rescaled to have mean 0 and variance equal to that in II-1. Barring the aforementioned changes, everything in study II remains the same as in I-1.

Study III evaluates the performance of RCRnorm in situations when samples are collected from different groups (or experimental conditions), and a subset of genes are differentially expressed (DE) between the groups. This can be done by randomly assigning each sample into one of the groups and then sampling $\kappa_{ir}$ from $N(\lambda_r + \Delta, \sigma_\kappa^2)$ for the samples from the treatment group and from $N(\lambda_r - \Delta, \sigma_\kappa^2)$ for those from the control group. If gene $r$ is DE, we set $\Delta = 0.2$; otherwise, $\Delta = 0$. Here, we assume two groups of equal size and set the proportion of DE genes to be 5%, 10% and 20% in settings III-1–III-3 respectively. Everything else remains the same as in I-1.

In study IV we generated raw counts for probe $j$ of sample $i$ from a negative binomial (NB) distribution NB($\xi_{ij}, \psi_j$) with mean $\xi_{ij}$ and variance $\xi_{ij} + \xi_{ij}^2/\psi_j$, where $\xi_{ij} > 0$, $\psi_j > 0$ and $j \in \{p, n, h, r\}$. With this mean-dispersion parameterization a small value of $\psi_j$ indicates large overdispersion of the observed count data while a large value of $\psi_j$ indicates that NB is approaching Poisson. We set $\psi_j \equiv 2$, 20 and 200 across all the probes in settings IV-1–IV-3 respectively. To generate the means, we employed a GLM $\log_{10} \xi_{ij} = a_i + b_i X_{ij} + d_j$, with $a_i$'s, $b_i$'s and $d_j$'s set to the same values as in I-1.

Lastly, we simulated data in the spirit of the NanoStringDiff model described in Wang et al. (2016) which is considerably different from the proposed model. To mimic the overdispersion observed in the real data, we replaced the Poisson kernel in the NanoStringDiff model with a NB kernel. For the regular genes we followed the notation used in Wang et al. (2016) and sampled $Y_{ir} \sim \text{NB}(c_i d_i \lambda_{ir} + \theta_{ir}, \psi)$, where $c_i$ is the positive control size factor, $d_i$ is the housekeeping size factor of the $i$th sample, $\lambda_{ir}$ is the unobserved expression rate and $\theta_{ir}$ is the nonspecific background noise of the $r$th regular gene from the $i$th sample. Read counts of negative controls, positive controls and housekeeping genes were generated in a similar manner but incorporated the characteristics of each control type. In settings V-1–V-3 we simulated the raw counts by setting the dispersion parameter $\psi = 2$, 20 and 200 respectively. All other parameter values were estimated from the lung cancer dataset we used.

Under each setting 50 datasets were independently simulated, each with 28 patient samples, six positive controls, eight negative controls, seven housekeeping genes and 83 regular genes which are exactly the same as in the real lung cancer data. Since all the read counts in studies I–III were generated in the log10 scale, they were exponentiated and rounded to the nearest integers so that the simulated data can be analyzed by all the algorithms. For RCRnorm 8000 iterations were simulated in each MCMC run, and the first 5000 were used for burn-in. The existing algorithms were applied using their default settings.

5.2. *Results.* To evaluate the performance on normalization under ground truth, we computed gene-wise correlations for 83 regular genes between normalized data and true expression levels within each simulated dataset and reported their mean, standard deviation (SD), 25th, 50th and 75th percentiles as the summary statistics for every method. Then, under each setting, boxplots of these summary statistics based on 50 replicates were generated and used to compare the four methods. Because RCRnorm works on the log10 scale, NAPPA on the log2 scale and the other algorithms on the original scale, we used Spearman correlation for comparison which ignores data values and relies solely on ranks.

Figure 2 shows boxplots of mean and SD, and Figure S2 in Supplementary Material (Jia et al. (2019)) shows boxplots of the 25th, 50th and 75th percentiles of gene-wise correlations for each of the five settings in Simulation I. In the basic setting I-1 RCRnorm is the winner while RUVseq performs the worst. The poor performance of RUVseq can be explained by the following reasons: (1) data from a medium throughput platform, such as nCounter, could not be well characterized by the model designed for high throughput sequencing data; (2) RUVSeq is problematic when there is only a small number of negative controls, as claimed by its authors. All other existing algorithms performed somewhat similarly, among which NanoStringDiff seems to fair a little worse than the other two methods, reflected by the generally smaller mean, percentiles, and larger SD of its correlations. In Setting I-2 the increased signal strength improves the performance of the existing algorithms significantly, but it does not affect RCRnorm much since RCRnorm already performs well in I-1. In settings I-3, I-4 and I-5, increased variability, regardless of the source, worsens the performance of every algorithm. Compared to I-1, the mean and percentiles decrease, but the SD increases in general. The increase in probe-level variation (i.e., $\sigma_d$ and $\sigma_{d-}$) in Setting I-4 has the largest negative impact on the performance, followed by the increase in variability of random noise (i.e., $\sigma_e$ and $\sigma_{e-}$) in Setting I-5 and last by the increase in lane-by-lane variation (i.e., $\sigma_a$ and $\sigma_b$) in Setting I-3. Among all, NanoStringDiff seems to be the most sensitive to such changes while RCRnorm is the least affected, maintaining a strong performance in all settings. In addition to being well apart from the other boxplots in all settings except for I-2, the boxplots for RCRnorm show the smallest interquartile ranges, meaning that RCRnorm gives very consistent results over different replicate datasets.

Figure 3, along with Figure S3 in Supplementary Material shows results for the six settings in Simulation II. With the gene-wise degradation levels added to the

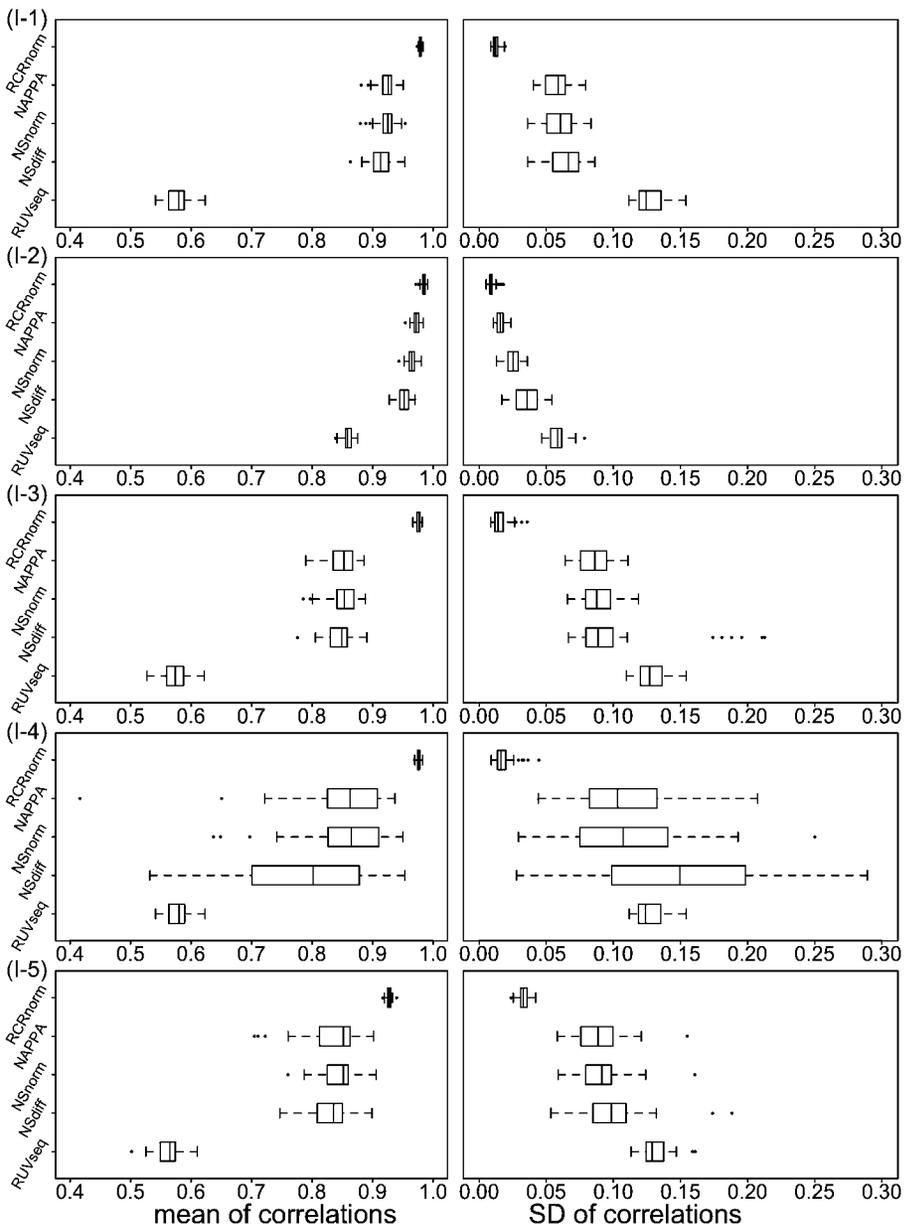

FIG. 2. *Simulation study* I *for mimicking various real-world scenarios*: *boxplots for mean and SD of gene-wise correlations between normalized data and true expression levels based on* 50 *replicates for each of the five settings* I1–I5. *Compared to the basic setting* I-1 (*parameter values estimated from the FFPE samples in the lung cancer application*), *gene expression variability is increased in* I-2, *lane-by-lane variation is increased in* I-3, *probe-level variation is increased in* I-4 *and variability of random noise is increased in* I-5. *Note that NSnorm stands for NanoStringNorm and NSdiff for NanoStringDiff.*

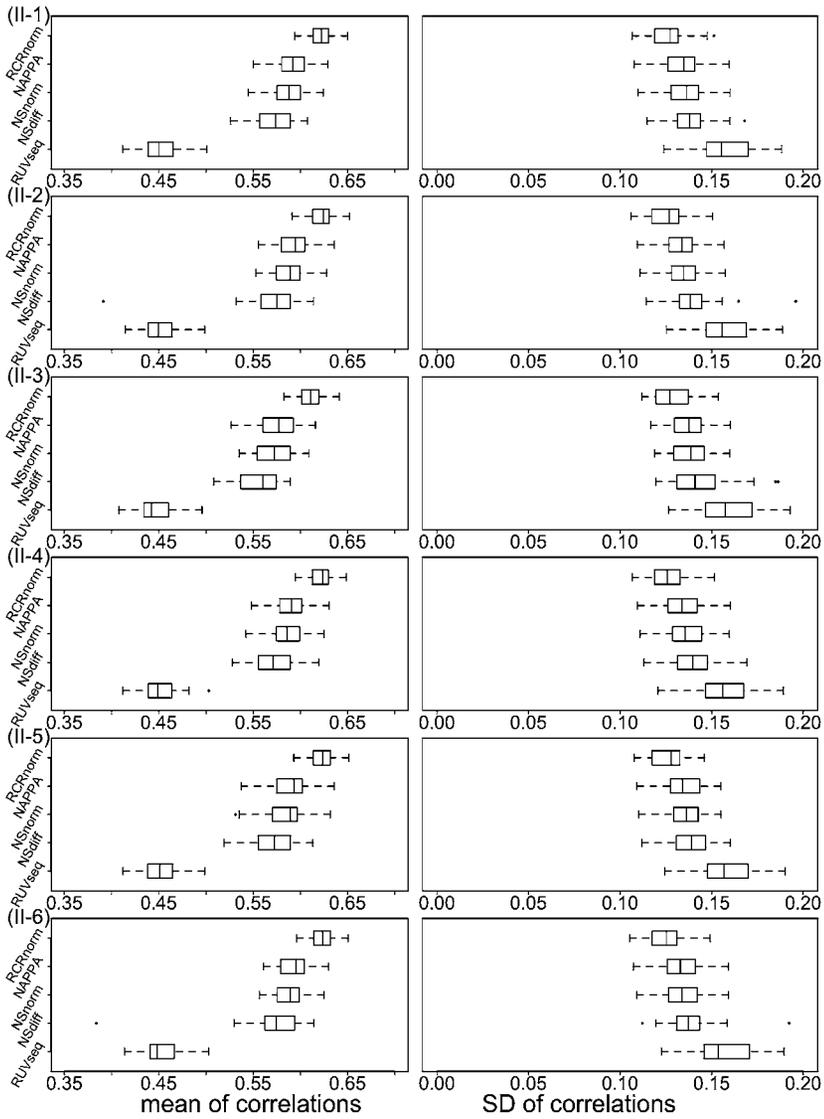

FIG. 3. *Simulation study* II *for mimicking different RNA degradation levels among genes* (II-1) *plus various distributional disturbances* (II-2–II-6): *boxplots for mean and SD of gene-wise correlations between normalized data and true expression levels based on* 50 *replicates for each of the six settings* II-1–II-6. *In* II-1 *gene-specific and sample-specific white noise* $N(0, \sigma_\delta^2 = 0.16)$ *was added to each common sample effect* $\phi_i$ *for all genes. In the other five settings distributional disturbances were further included: in* II-2 *a thick-tail distribution* $t_3$ *was used to simulate probe effects; in* II-3 $t_3$ *was used to simulate random errors; in* II-4 *a right-skewed distribution* $G_{2,1}$ *was used to simulate probe effects; in* II-5 $G_{2,1}$ *was used to simulate random errors, and in* II-6 $t_3$ *was used to generate probe effects and* $G_{2,1}$ *to generate random errors. Except for the changes above, everything remains the same as in the basic setting* I-1, *including all parameter values (so* $t_3$ *and* $G_{2,1}$ *need to be rescaled or shifted).*

model system, the performance of every algorithm in II-1 becomes (much) worse than the performance in I-1, as indicated by the generally wider boxes for all the summary statistics and the lower box centers for the mean and percentiles. By contrast the distributional disturbances seem to not affect the performance much. The differences between II-1 and the other five II settings are not large when compared to those between I-1 and II-1. Among these disturbances heavy-tailed random errors in II-3 tend to make the methods perform worse than the other disturbances. Overall, RCRnorm is quite robust to the moderate violations of the normality assumption and outperforms the other methods.

Figure 4 and Figure S4 in Supplementary Material show that when the proportion of DE genes is small, which is typical in practical situations, increasing the DE proportion would not affect the performance much for any of the five methods. Figures 5 and 6, along with Figures S5 and S6 in Supplementary Material show results for count data generated from the NB GLM setup and the NanoStringDiff model. Clearly, when the overdispersion of count data is large (i.e., small $\psi$), the performance becomes much worse, compared to that for larger $\psi$ values; among all, NanoStringDiff is the most sensitive as it relies on the Poisson assumption. In general, RCRnorm maintains the best performance in these studies, though its advantage over the other four methods weakens in Study V where the proposed models in Section 3 no longer hold.

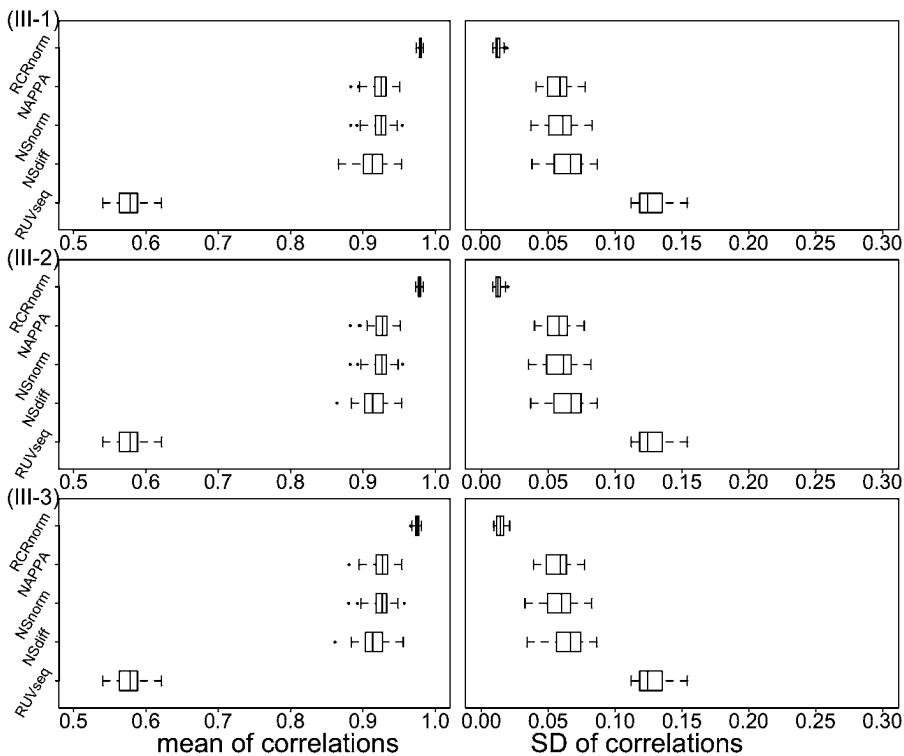

FIG. 4. *Simulation study* III *for mimicking situations when there is a subset of regular genes that are differentially expressed between different groups: boxplots for mean and SD of gene-wise correlations between normalized data and true expression levels based on* 50 *replicates for each of the three settings* III-1–III-3 *which correspond to different DE proportions* 5%, 10% *and* 20% *respectively.*

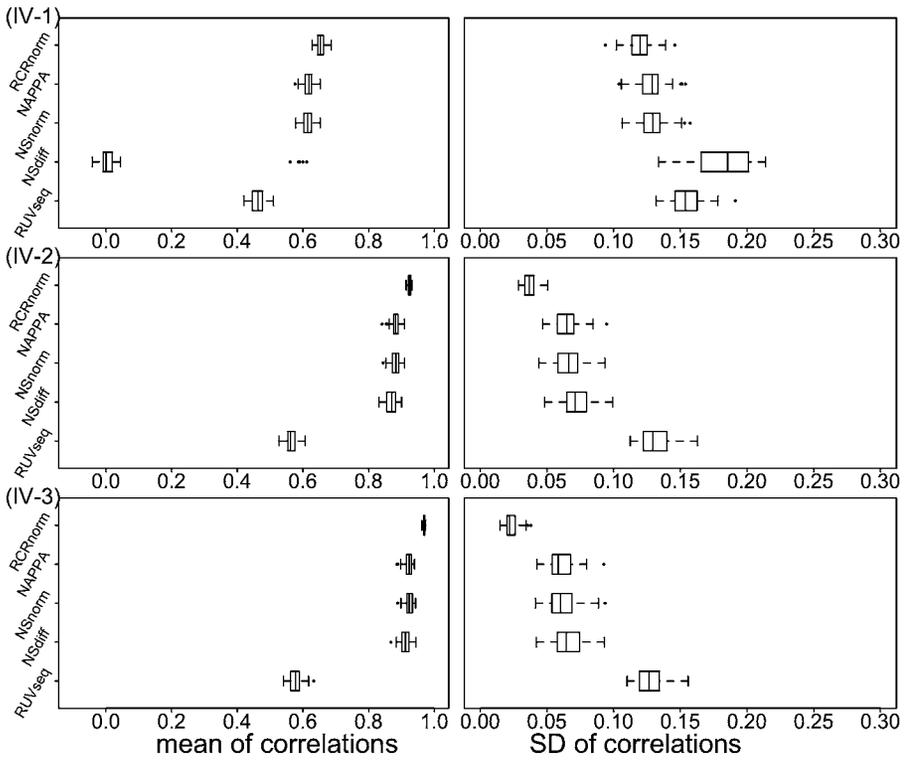

FIG. 5. *Simulation study* IV *for generating the raw counts from negative binomial (NB) GLMs: boxplots for mean and SD of gene-wise correlations between normalized data and true expression levels based on* 50 *replicates for each of the three settings* IV-1–IV-3 *which correspond to different dispersion parameters* $\psi = 2$, 20 *and* 200 *respectively. Note that, as* $\psi$ *increases, NB approaches Poisson.*

## 6. Real data applications.

6.1. *Lung cancer data.* We use the NanoString nCounter data from FFPE samples described in Section 2 to illustrate the proposed RCRnorm first. We ran our MCMC algorithm for 15,000 iterations in total. The convergence for all the model parameters was detected after 7000 iterations, and we discarded the first 10,000 for burn-in. We then thinned the chain to reduce the autocorrelation among posterior draws by saving every tenth draw only, so, in total, 500 posterior samples were kept. Figure 7 shows the posterior densities of global parameters includ-ing $\mu_a$, $\mu_b$, $c$, $\sigma_{\kappa*}$, $\sigma_\kappa$, $\sigma_d$, $\sigma_{d-}$, $\sigma_e$ and $\sigma_{e-}$, and Table 1 presents a summary for Bayesian estimates of these parameters including the posterior mean, median, standard error (SE), and a 95% credible interval (CI) using the 2.5th and 97.5th percentiles of (thinned) posterior samples. Here, the posterior mean is used to estimate each location parameter, but the posterior median is used to estimate each variance parameter since the corresponding posterior density is skewed to the right.

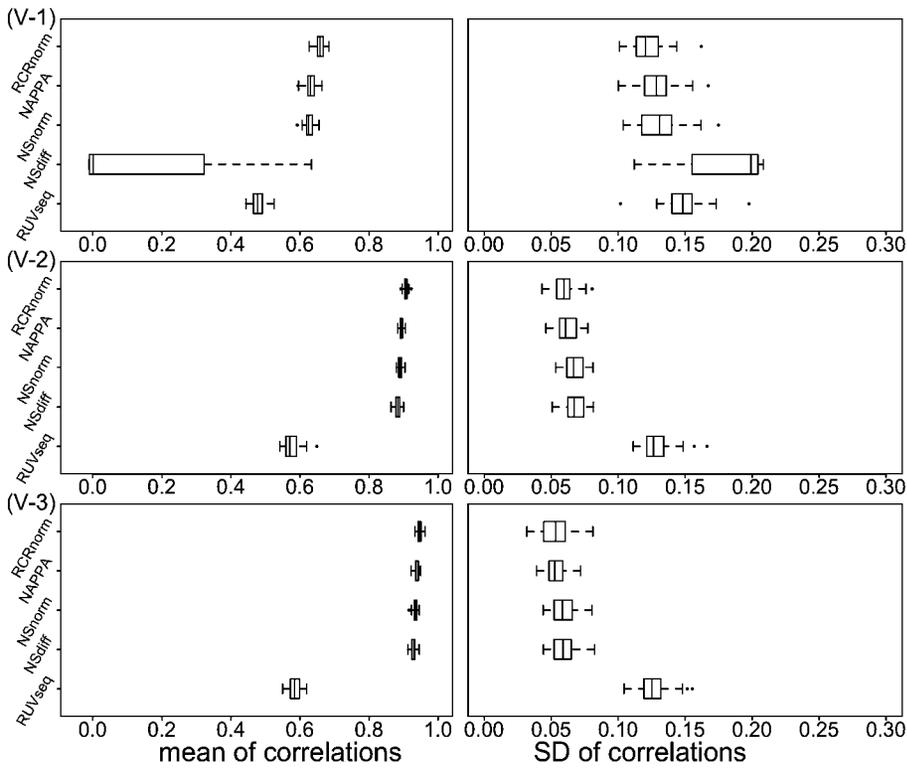

FIG. 6. *Simulation study* V *for generating the raw counts from the NanoStringDiff model in* Wang et al. (2016) *with the Poisson kernel replaced by the negative binomial (NB) kernel—boxplots for mean and SD of gene-wise correlations between normalized data and true expression levels based on* 50 *replicates for each of the three settings which correspond to different dispersion parameters* $\psi = 2$, 20 *and* 200 *respectively. Note that as* $\psi$ *increases, NB approaches Poisson.*

Several intriguing observations can be made from the above figure and table. As we know, housekeeping genes are involved in the maintenance of basic cellular function, so they are expected to be uniformly expressed with low variability in all cells and experimental conditions. Our analysis using RCRnorm confirms that, compared to other genes, expression levels of housekeeping genes indeed have much less variation. Clearly, the Bayesian estimate of $\sigma_{\kappa*}$ (0.136, the SD of expression levels for housekeeping genes) is much smaller than that of $\sigma_\kappa$ (0.361, the SD of expression levels for regular genes), and Figure 7(b) shows that their

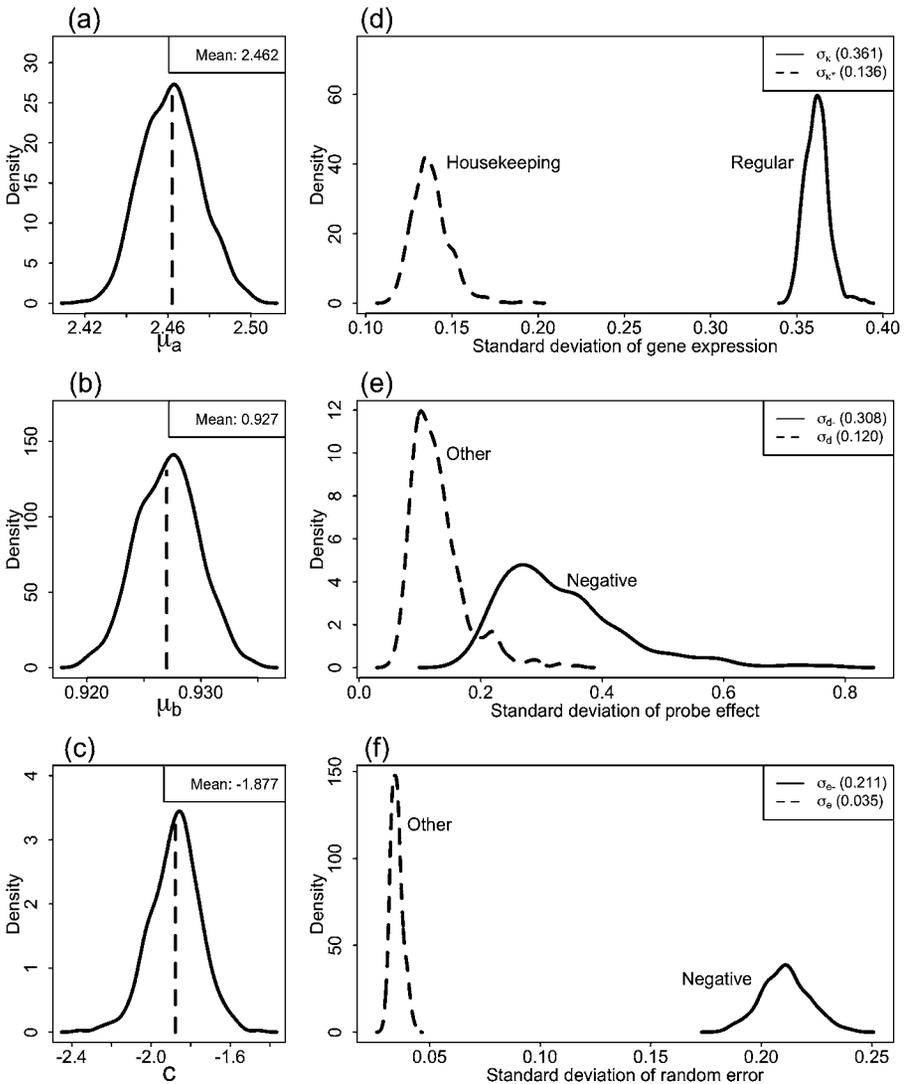

FIG. 7. *Lung cancer data example: posterior densities of global parameters from applying RCRnorm to FFPE samples, where panel* (a) *is for* $\mu_a$, (b) *for* $\mu_b$, (c) *for* $c$, (d) *for* $\sigma_\kappa$ *and* $\sigma_{\kappa*}$, (e) *for* $\sigma_d$ *and* $\sigma_{d-}$ *and* (f) *for* $\sigma_e$ *and* $\sigma_{e-}$ *respectively.*

posterior density curves are well separated with the correct order. Furthermore, our exploratory analysis in Section 2.2 strongly indicates $\sigma_d < \sigma_{d-}$ and $\sigma_e < \sigma_{e-}$. That is, the SD of the probe-specific deviation (from the linear trend) for negative controls is larger than that for the other types of probes, as is the SD of the random errors. The Bayesian estimates in Table 1 confirm the underlying features again $(0.120 < 0.308$ and $0.035 < 0.211)$, and the posterior densities in Figures 7(e) and

TABLE 1
*Lung cancer data example: posterior summary statistics of global parameters from applying RCRnorm to FFPE samples*

|  | Mean | Median | SD | 95% CI |
|---|---|---|---|---|
| $\mu_a$ | 2.462 | 2.462 | 0.014 | (2.436, 2.490) |
| $\mu_b$ | 0.927 | 0.927 | 0.003 | (0.922, 0.932) |
| $c$ | $-1.877$ | $-1.872$ | 0.125 | $(-2.126, -1.635)$ |
| $\sigma_{\kappa*}$ | 0.138 | 0.136 | 0.011 | (0.121, 0.163) |
| $\sigma_\kappa$ | 0.361 | 0.361 | 0.007 | (0.349, 0.375) |
| $\sigma_{d-}$ | 0.333 | 0.308 | 0.104 | (0.198, 0.595) |
| $\sigma_d$ | 0.129 | 0.120 | 0.045 | (0.074, 0.242) |
| $\sigma_{e-}$ | 0.211 | 0.211 | 0.011 | (0.189, 0.233) |
| $\sigma_e$ | 0.035 | 0.035 | 0.003 | (0.031, 0.041) |

(f) support them as well. Note that the data model of RCRnorm does not impose such order constraints at all, but the end results from RCRnorm capture these characteristics accurately.

Next, we compare the performance of RCRnorm in normalizing FFPE data with the existing algorithms. As FF samples generally have much better quality than FFPE samples, we used normalized FF data as the gold standard, where each method was applied to normalize both FFPE and FF data, and Pearson correlation coefficients between normalized FFPE and FF data were computed to quantify its performance. The summary statistics (mean, SD, 25%, 50% and 75% quantile) of 83 gene-wise correlations are presented in the left panel of Table 2 with the best value bolded in each column. Compared to the original data, all algorithms except for RUVseq (designed for high-throughput RNA-seq data) significantly improve the gene-wise correlations, and RCRnorm has the best performance in terms of higher mean and percentiles as well as smaller variability. Note that the three existing algorithms designed for the NanoString nCounter platform have somewhat similar performance. However, RCRnorm can further offer a sizable gain over their already improved performance.

Although gene-wise correlations are the focus of the original study (Xie et al. (2017)) for validating the gene signature, we also report 28 patient-wise correlations in the right panel of Table 2. Here, NanoStringDiff and NanoStringNorm are unable to adjust patient-wise correlations. This is because they linearly transform a patient's data with scale factors calculated from internal controls and, thus, do not change the patient-wise correlations. Among the other three methods RCRnorm and NAPPA can achieve some improvement from normalization, though improvement is not as large as seen in the gene-wise case. This is perhaps because the patient-wise correlations of the original data are already high, leaving little room for improvement. Again, RCRnorm seems to be better than NAPPA with higher mean, percentiles and similar variability.

*Lung cancer data example: summary statistics of gene-wise and patient-wise correlations between normalized FFPE and FF samples using different algorithms*

| | Gene-wise correlation | | | | | Patient-wise correlation | | | | |
|---|---|---|---|---|---|---|---|---|---|---|
| | Mean | SD | 25% | 50% | 75% | Mean | SD | 25% | 50% | 75% |
| Original data | 0.291 | 0.278 | 0.114 | 0.303 | 0.444 | 0.806 | 0.129 | 0.720 | 0.835 | 0.897 |
| RCRnorm | **0.550** | 0.197 | **0.427** | **0.590** | **0.694** | **0.851** | **0.102** | 0.792 | **0.876** | **0.926** |
| NAPPA | 0.488 | **0.194** | 0.370 | 0.522 | 0.646 | 0.844 | **0.102** | **0.793** | 0.863 | 0.919 |
| NanoStringDiff | 0.487 | 0.207 | 0.352 | 0.496 | 0.635 | 0.806 | 0.129 | 0.720 | 0.835 | 0.897 |
| NanoStringNorm | 0.489 | 0.199 | 0.341 | 0.474 | 0.630 | 0.806 | 0.129 | 0.720 | 0.835 | 0.897 |
| RUVseq | 0.300 | 0.276 | 0.103 | 0.308 | 0.465 | 0.806 | 0.118 | 0.721 | 0.819 | 0.904 |

For the unnormalized data the patient-wise correlations (mean: 0.806, SD: 0.129) are much higher than the gene-wise correlations (mean: 0.291, SD: 0.278). While a few highly expressed genes, if any, could inflate their values, such higher patient-wise correlations may still suggest the superb performance of NanoString nCounter on expression profiling with FFPE samples. On the other hand the much lower gene-wise correlations highlight the importance of removing sample-specific effects for downstream statistical analysis.

In this application the gene-wise correlations achieved by all the algorithms are generally lower than what we have seen in our simulation. This can be explained by the following reasons. First, due to experimental and technical limitations, gene-wise RNA degradation levels cannot be measured and removed by any of the five algorithms. As shown in Figure 3, such variation could cause a large drop in performance. Second, normalized FF expression levels were used to calculate the Pearson correlation coefficients in the application while true expression levels were used in the simulation.

6.2. *Colorectal cancer data.* Our second application involves a colorectal cancer study (Omolo et al. (2016)) that compared five different platforms to identify which platform could faithfully translate the RAS pathway gene signature identified from FF samples into FFPE samples. RAS pathway activation is a risk factor for the failure of EGFR combination therapy in colorectal cancer patients. Thus, it is clinically important to identify a platform that can obtain reliable information from FFPE samples. Among the five platforms compared, NanoString nCounter was found to be the best platform to recover gene expression information from FFPE samples.

We applied RCRnorm to the NanoString nCounter data from FFPE samples in this study where, again, 15,000 iterations were used with the first 10,000 being burn in. The dataset contains 54 samples, with six positive controls (with the same input amounts as before), six negative controls, 11 housekeeping genes and 18

|  | RCRnorm | NAPPA | NanoStringDiff | NanoStringNorm | RUVseq |
| --- | --- | --- | --- | --- | --- |
| $p$ value | 0.036 | 0.146 | 0.170 | 0.096 | 0.004 |

regular genes. Thus, unlike the lung cancer data, this dataset has more samples than probes.

Figure S7 plots posterior densities, and Table S2 reports posterior summary statistics of global parameters in Supplementary Material (Jia et al. (2019)). Compared with Table 1 in the lung cancer study, the estimates of $\mu_a$, $\mu_b$ and $c$ are all close, although they are completely from two independent and distinct studies. Again, the Bayesian estimates of $\sigma_{\kappa*}$ and $\sigma_\kappa$ (0.190 vs. 0.278) confirm that expression levels of the housekeeping genes vary less than those of the regular genes. Interestingly, $\sigma_{\kappa*}$ in this study is larger than its counterpart in the lung cancer study (estimate: $0.190 > 0.136$; 95% CI: $(0.176, 0.215)$ completely above $(0.121, 0.163)$), even though more housekeeping genes were used here (11 vs. 7). This seems to suggest that increasing the number of housekeeping genes used does not necessarily reduce their variability to the minimal level. Our results further confirm that $\sigma_d \ll \sigma_{d-}$ (0.069 vs. 0.257) and $\sigma_e \ll \sigma_{e-}$ (0.042 vs. 0.273). Recall that these characteristics were revealed by exploring the lung cancer data. Nevertheless, they may generally hold for NanoString nCounter data of any kind.

In this study paired FF NanoString data are not available, and so performance comparison among the five methods cannot be done using correlations between normalized FFPE and FF data. According to the manual of NanoStringNorm (Waggott et al. (2012)), housekeeping genes are typically selected to be genes with high means and low standard deviations. Thus, the two-sample $t$-test was used to compare the coefficients of variation (CV = SD/mean) between the housekeeping and regular genes. A good normalization algorithm should have a clear separation of CVs between these two types of genes. To make the CVs from the five algorithms comparable, we transformed the normalized data from the algorithms into the same scale. In Table 3 RCRnorm and RUVseq have $p$-values smaller than 0.05 which confirms the existence of a significant difference between the two gene types at a significance level of 0.05. None of the existing NanoString methods was able to do so. Also, RUVseq performed poorly in simulation when the ground truth is known and, in the lung cancer study, when paired FF data are available. Putting these together, we conclude that RCRnorm has consistently strong performance in normalizing NanoString data.

**7. Discussion.** Motivated by a lung cancer study in predicting adjuvant chemotherapy (ACT) response, we have developed a novel (Bayesian) method,

RCRnorm, to normalize NanoString nCounter data. Through simulation studies, we have shown that RCRnorm compares very favorably with the existing methods, especially in situations with an elevated level of heterogeneity from various sources. In the lung cancer application RCRnorm performs the best and greatly improves gene-wise correlations between paired FF and FFPE samples. Thus, it provides an important step toward applying gene signatures identified from genome-wide expression profiling of FF samples into wide clinical use with FFPE samples. In addition RCRnorm provides improved patient-wise correlations that NanoStringDiff and NanoStringNorm cannot.

The competitive performance of RCRnorm can be largely explained by two unique features it owns. First of all, RCRnorm relies on an integrated system of hierarchical linear regression models with random coefficients, which effectively captures mean and variance structures underlying the data shared among different types of probes, to maximally remove systematic sample-specific biases in gene expression profiling. Unlike RCRnorm, the previous methods use the different types of internal controls in an isolated and somewhat heuristic manner. Thus, they do not take full advantage of the rich information provided by the nCounter system. Second, the existing methods adjust sample loading effects with information extracted from housekeeping genes. This is based on the assumption that expression levels of housekeeping genes are stable across samples, implying that their biological variability in gene expression is zero. In practice their biological variability, although smaller than other genes in general, is not zero. In fact some housekeeping genes have been reported to have significant fluctuations (Gubern et al. (2009)). Thus, this simplifying assumption may lead to compromised performance of the existing methods on normalization. By contrast RCRnorm does not need the assumption as two separate variance terms, $\sigma_{\kappa*}^2$ and $\sigma_{\kappa}^2$, are used to model the biological variability of housekeeping and regular genes respectively. Further, by estimating and comparing $\sigma_{\kappa*}^2$ and $\sigma_{\kappa}^2$, RCRnorm can provide an alternative way to examine the validity of housekeeping genes used in an nCounter system from an analytical perspective.

In addition RCRnorm offers much better interpretability than the other existing methods. It is based on a rigorous model system whose parameters can be intuitively interpreted. With estimates obtained from the Bayesian approach, researchers can gain a deep understanding about the dataset under study. Moreover, in the integrated system, $X$ represents the log RNA content whose value is attached with a unit log fM; so, for housekeeping and regular genes, the quantity of interest $\kappa$, which is an additive term that makes up $X$, should have the same unit. Thus, the normalized expression produced by RCRnorm also has the unit log fM so that its values are directly comparable to the input amounts of positive controls in the system. This may help us concretely understand the level of expression for a particular gene. It is also worth mentioning that RCRnorm does not depend on the log transformation at all, and the response $Y$ in equations (2)–(5) can be $g$-transformed from raw read counts where $g(\cdot)$ can be any function such that $Y$ can be approximately modeled by normal distributions. Therefore, for a real data set, we can use exploratory analysis to decide which transformation works better and then preprocess the data accordingly before applying RCRnorm.

RCRnorm employs a Bayesian framework to handle its computational needs, so a noninformative prior setup, as detailed in Section 4.2, is adopted. In situations where meaningful prior knowledge is available, prior distributions can be chosen to incorporate such knowledge for improved results. We have developed an efficient Gibbs sampler for posterior computation and inference, where all the steps can be done by direct sampling from known distributions. Based on our numerical experience, the algorithm is computationally stable and converges without manual tuning in all our settings. In the lung cancer application it took ∼1000 seconds to run 15,000 iterations using R on a 2.8 GHz Intel Core i7 processor. Note that, based on the proposed model system, frequentist approaches to estimation such as maximum likelihood and nonparametric methods may be used to facilitate computation. Nevertheless, RCRnorm offers the advantage of quantifying the estimation uncertainty easily as it is Bayesian in nature. The existing algorithms, which are all frequentist, cannot even provide confidence intervals for key parameters or normalized expression.

RCRnorm is a method designed based on characteristics of nCounter data observed from FFPE samples. When developing normalization methods, the focus on FFPE samples itself is original as no previous research has even realized that FFPE data are more complex than FF data. On the other hand this focus does *not* limit the application of RCRnorm to FF or like samples, as the approach that starts with a complex situation and then deduces a simpler case is valid. RCRnorm can also be used with other platforms that use internal controls (with minor adaptions). Furthermore, replicates from the same patient, when available, can be naturally incorporated into the system to enable better estimation of model parameters and normalized expression. This is because RCRnorm, again as a Bayesian method, has the capability to pool information from various sources such as probes, patients or replicates. The other methods can only treat the replicates as independent samples and then calculate the mean or median of normalized expression levels to combine replicates.

Although current technologies cannot reliably quantify RNA degradation levels in FFPE samples, clinical information for these samples such as age of FFPE samples (not the patient age) and RIN (RNA integrity number; a measure used to evaluate RNA quality) (von Ahlfen et al. (2007)), if available, can be integrated to model the degradation levels to potentially improve the performance on normalization.

As mentioned in Vallejos et al. (2017), there are two different approaches to normalization. The first is to use generic methods that yield normalized expression measures that can be used as input in any downstream analysis; the second is to consider *bespoke* methods that use prenormalized expression measures in conjunction with a model that can account for both effects of interest and unwanted artifacts. Like most existing normalization methods, RCRnorm belongs to the first approach. If there are two groups of samples (say treatment and control), the treatment effect, if any, should be harbored in the $\kappa_{ir}$ terms (i.e., normalized expressions). However, in such situations, a *bespoke* method should be more efficient than a generic method because: (i) it can directly model the group label, and (ii) it can integrate the normalization step into the differential expression (DE) analysis so that a two-stage approach that often ignores the uncertainty from the first stage is no longer needed. Thus, developing a *bespoke* method based on some modifications of RCRnorm for DE analysis of NanoString nCounter data would be an interesting topic for future research.

## SUPPLEMENTARY MATERIAL

**Supplement to "RCRnorm: An integrated system of random-coefficient hierarchical regression models for normalizing NanoString nCounter data"** (DOI: 10.1214/19-AOAS1249SUPP; .pdf). Supplement A: tables, figures and Bayesian full conditionals. In the first section of this Supplementary Material, we provide additional tables and figures mentioned in the paper, including additional information about lung cancer data, additional simulation results, and results for colorectal cancer data. In the second section of this Supplementary Material, we provide the full posterior conditionals for the Gibbs sampler used to draw posterior samples.

G. JIA
DEPARTMENT OF STATISTICAL SCIENCE
SOUTHERN METHODIST UNIVERSITY
3225 DANIEL AVENUE
P.O. BOX 750332
DALLAS, TEXAS 75275
USA
AND
QUANTITATIVE BIOMEDICAL RESEARCH CENTER
DEPARTMENT OF CLINICAL SCIENCES
UNIVERSITY OF TEXAS SOUTHWESTERN
   MEDICAL CENTER
DALLAS, TEXAS 75390
USA

X. WANG
DEPARTMENT OF STATISTICAL SCIENCE
SOUTHERN METHODIST UNIVERSITY
3225 DANIEL AVENUE
P.O. BOX 750332
DALLAS, TEXAS 75275
USA
E-MAIL: swang@smu.edu

Q. Li
Y. Xie
Quantitative Biomedical Research Center
Department of Clinical Sciences
University of Texas Southwestern
    Medical Center
Dallas, Texas 75390
USA

W. Lu
X. Tang
I. Wistuba
MD Anderson Cancer Center
Department of Translational
    Molecular Pathology
University of Texas
Houston, Texas 77030
USA